\documentclass[aps,prl,showpacs,twocolumn,a4paper]{revtex4}
\usepackage{graphicx}
\usepackage{amsmath}
\usepackage{amsfonts}

\begin{document}

\title{Four-Step Evolution of Spin-Hall Conductance: Tight-Binding
Electrons with Rashba Coupling in a Magnetic Field}
\author{Yi-Fei Wang$^{1}$, Yang Zhao$^{1}$, and Chang-De Gong$^{2,1}$}
\affiliation{$^1$National Laboratory of Solid State Microstructures and
Department of Physics, Nanjing University, Nanjing 210093, China\\
$^2$Department of Physics, Suzhou University, Suzhou 215006, China}
\date{\today}

\begin{abstract}
An intriguing magneto-transport property is
demonstrated by tight-binding lattice electrons with Rashba
spin-orbit coupling (SOC) in a magnetic field. With the flux
strength $\phi={2\pi/N}$ ($N$ is an integer) and the Zeeman
splitting fixed, when increasing the Rashba SOC $\lambda$,
the spin-Hall and
charge-Hall conductances (SHC and CHC) undergo four-step evolutions: the SHC shows
size-dependent resonances and jumps at three critical $\lambda_{c}$'s, and changes
its sign at $\lambda_{c1}$ and $\lambda_{c3}$; while the CHC exhibits three
quantum jumps by $-Ne^2/h$, $+2Ne^2/h$ and
$-Ne^2/h$. Such four-step evolutions are also
reflected in topological characters and spin polarizations of edge states of a cylindrical system, and are robust
against weak disorder.

\end{abstract}

 \maketitle

{\it Introduction.---}Recently, the spin-Hall effect (SHE), i.e., a
generation of spin current perpendicular to an applied electric
field~\cite{Dyakonov,Murakami1,Sinova,Kato}, has shed new light on
spintronics~\cite{Wolf} and provided novel techniques to manipulate
spins in nanostructures. In contrast to the extrinsic
SHE driven by spin-orbit (SO) impurity scattering~\cite{Dyakonov}, it is proposed that an intrinsic SHE exist
in semiconductors with SO coupled
bands~\cite{Murakami1,Sinova}. These proposals encouraged
the discovery of the SHE in GaAs semiconductor films
and heterostructures~\cite{Kato}, and in metallic Al films and Pt strips~\cite{Valenzuela}. In models with SO
coupled bands, two-dimensional electron gas (2DEG) with Rashba
spin-orbit coupling (SOC)~\cite{Rashba1} has the simplest form and
is therefore most notable~\cite{Sinova,Sheng2,Rashba2,Engel,Shen}.
Meanwhile, tunable Rashba SOC has been achieved via an external gate
voltage on the top of asymmetric heterostructures~\cite{Nitta}, and the Rashba SO field in quantum wells and semiconductors can
also be measured optically~\cite{Meier}.

In a clean 2DEG with parabolic dispersion and linear Rashba SOC, Sinova {\it et al.}
predicted that the spin-Hall conductance (SHC) holds a universal value
independent of SOC strength when both SO split bands are
occupied~\cite{Sinova}. It is now known that such an
intrinsic SHC with only linear Rashba SOC might
be destroyed by any amount of disorder~\cite{Engel}, or be canceled
completely by intrabranch contributions in the presence of a magnetic
flux~\cite{Rashba2}. In parallel, the SHC of 2DEG with linear Rashba
SOC and Zeeman splitting in a magnetic field was
calculated, and a resonant SHC was predicted when two Landau levels
cross each other at the Fermi level~\cite{Shen}.

In the presence of an underlying lattice potential, e.g. in metallic
conductors like Al films and Pt strips~\cite{Valenzuela},
both parabolic dispersion and linear SOC should be modified and
then incorporated into a lattice model which has been employed
to study the effect of disorder on the SHE in the metallic regime~\cite{Sheng2}.
Here, we begin to investigate a lattice system of 2D tight-binding electrons (TBE)
with Rashba SOC in a magnetic field.
This model is also relevant to novel experimental systems such as
untracold fermions in an optical
lattice with an effective SOC~\cite{Stanescu}
and graphene with an intrinsic or Rashba SOC~\cite{Kane,Sheng3}.
We focus on magneto-transport properties, and have found that
tuning the Rashba SOC strength generates novel four-step evolutions
of the SHC and the charge-Hall conductance (CHC).
Such bulk properties are also
reflected in topological characters and spin polarizations of edge states
of a cylindrical system, and are robust against weak disorder.

{\it Formulation.---}The model Hamiltonian of 2D TBE on a
square lattice with Rashba SOC and a uniform perpendicular magnetic
field $\vec{B}=(0,0,-B)$ is~\cite{Sheng2}:
\begin{eqnarray}
H&=&-t\sum_{\langle{ij}\rangle}\left[e^{i\phi_{ij}}\hat{c}^{\dagger}_{i}\hat{c}_{j}+\text{H.c.}\right]
+\lambda\sum_{i}\left[ie^{i\phi_{i,i+\vec{y}}}\hat{c}^{\dagger}_{i}\sigma_{x}\hat{c}_{i+\vec{y}}\right.\nonumber\\
& &\left.
-ie^{i\phi_{i,i+\vec{x}}}\hat{c}^{\dagger}_{i}\sigma_{y}\hat{c}_{i+\vec{x}}+\text{H.c.}\right]
-h_{z}\sum_{i}\left(n_{i\uparrow}-n_{i\downarrow}\right) \label{e.1}
\end{eqnarray}
where
$\hat{c}^{\dagger}_{i}=(c^{\dagger}_{i\uparrow},c^{\dagger}_{i\downarrow})$
are electron creation operators at site $i$, $\sigma_{x}$ and
$\sigma_{y}$ are Pauli matrices, the nearest-neighbor hopping
integral $t$ will be taken as the unit of energy, $\lambda$ is the
Rashba SOC strength, and the Zeeman splitting parameter is
$h_{z}={1\over2}g\mu_{b}B$ with $g$ the Land\'{e} factor and
$\mu_{b}$ the Bohr magneton. The magnetic flux per plaquette
is $\phi=\sum_{\square}\phi_{ij}=2\pi Ba^2/\phi_0={2\pi/N}$
with $N$ an integer, $a$ the lattice constant and $\phi_0=hc/e$ the
flux quantum. The Landau gauge $\vec{A}=(0,-Bx,0)$ and the
corresponding periodical boundary conditions (PBCs) are adopted, and
the magnetic unit cell has the size $N\times 1$.

After the numerical diagonalization of the Hamiltonian
[Eq.~(\ref{e.1})], the zero-temperature ($T=0$) CHC is calculated
through the Kubo formula~\cite{Thouless}
\begin{eqnarray}
&&{\sigma}_{\rm CH}(E)={i{e^2\hbar}\over{A}}\sum_{\varepsilon_{m{\bf
k}}<E}\sum_{\varepsilon_{n{\bf k}}>E}\nonumber\\
&&{{{\langle m{\bf k}|v_x|n{\bf k}\rangle\langle n{\bf k}|v_y|m{\bf
k}\rangle -\langle m{\bf k}|v_y|n{\bf k}\rangle\langle n{\bf
k}|v_x|m{\bf k}\rangle}}\over{{(\varepsilon_{m{\bf
k}}-\varepsilon_{n{\bf k}})}^2}}\label{e.2}
\end{eqnarray}
while the SHC at $T=0$ is given by~\cite{Sinova}
\begin{equation}
{\sigma}_{\rm SH}(E)=-{{e\hbar}\over{A}}\sum_{\varepsilon_{m{\bf
k}}<E}\sum_{\varepsilon_{n{\bf k}}>E}{\text{Im}{\langle m{\bf
k}|J^{z\text{spin}}_{x}|n{\bf k}\rangle\langle n{\bf k}|v_y|m{\bf
k}\rangle}\over{{(\varepsilon_{m{\bf k}}-\varepsilon_{n{\bf
k}})}^2}} \label{e.3}
\end{equation}
where $A=L\times L$ is the area of this 2D system, $E$ is the Fermi
energy, $\varepsilon_{m{\bf k}}$ is the corresponding eigenvalue of
the eigenstate $|m{\bf k}\rangle$ of $m$th Landau subband, and the
summation over wave vector ${\bf k}$ is restricted to the magnetic
Brillouin zone (MBZ): $-\pi/N\leq{k_x}a<\pi/N$ and
$-\pi\leq{k_y}a<\pi$. The velocity operator is defined as ${\bf
v}=i/\hbar[H,{\bf R}]$ ({\bf R} is the position operator of
electron) and the spin current operator as
$J^{z\text{spin}}_{x}=\hbar/4\{v_{x},\sigma_{z}\}$. When $E$ falling
in energy gaps, we can rewrite $\sigma_{\rm CH}$ as $\sigma_{\rm
CH}(E)=e^2/h\sum_{\varepsilon_m<E}C_m$, where $C_m$ is the Chern
number~\cite{Thouless} of the $m$th totally filled Landau subband.

\begin{figure}[!htb]
  \vspace{-0.25in}
  \hspace{-0.18in}
  \includegraphics[scale=0.8]{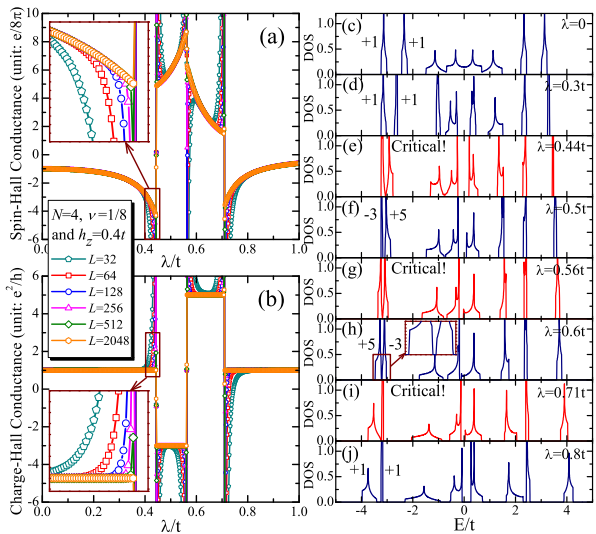}
  \vspace{-0.45in}
  \caption{(color online). The case with $N=4$ and $h=0.4t$.
  (a) The spin-Hall conductance $\sigma_{\rm SH}$ versus the Rashba SOC parameter $\lambda$ for electron filling $\nu={1\over{8}}$
  and various lattice sizes. (b) The charge-Hall conductance $\sigma_{\rm CH}$ versus $\lambda$
  in the cases of (a). (c)-(j) The DOS for some $\lambda$'s in (a). The Chern numbers of subbands are also shown.} \label{f.1}
\end{figure}

{\it An example with $N=4$.---}An overall picture of the
CHC $\sigma_{\rm CH}$ and the SHC $\sigma_{\rm SH}$ calculated by Eq.~(\ref{e.2}) and Eq.~(\ref{e.3})
are shown in Fig.~\ref{f.1} with $N=4$ (i.e., the flux strength
$\phi={1\over{4}}\times{2\pi}$), $h_{z}=0.4t$ and various lattice
sizes with $L=32-2048$. We concentrate on the lowest
Landau subbands and consider the electron filling $\nu={1\over{8}}$.

In the case of $\lambda=0$
[Fig.~\ref{f.1}(c)], the density of states (DOS) is symmetric about
the Fermi energy $E$, and the lowest two Landau subbands (each
totally-filled subband contributes ${1\over{8}}$ to $\nu$) are well
separated, each carrying a Chern number $+1$. With $\lambda$
increasing from $0$ to $1.0t$ one sees a systematic four-step
evolution of $\sigma_{\rm CH}$ and $\sigma_{\rm SH}$ versus
$\lambda$; there are three critical $\lambda_{c}$'s at which both
$\sigma_{\rm SH}$ and $\sigma_{\rm CH}$ exhibit jumps.

When $\lambda$ increases from $0$ to $\lambda_{c1}\approx0.44t$, the
lowest two Landau subbands approach each other, then merge together
and form a pseudogap at $\lambda_{c1}$ [Fig.~\ref{f.1}(e)];
$\sigma_{\rm SH}$ changes continuously from $-1{e/{8\pi}}$ to larger
negative values [Fig.~\ref{f.1}(a)]; while $\sigma_{\rm CH}=+1e^2/h$
nearly stays unchanged [Fig.~\ref{f.1}(b)]. Here for a small
lattice size ($L=32$), $\sigma_{\rm SH}$ and $\sigma_{\rm CH}$
both present divergence when $\lambda$ approaches $\lambda_{c1}$.
With the lattice size increased ($L=64-512$), the divergence is
weakened accordingly; for $L=2048$, $\sigma_{\rm SH}$ approaches a finite value
$-4.30{e/{8\pi}}$ at $\lambda_{c1}$, and $\sigma_{\rm CH}$ remains
as $+1e^2/h$ for $0\leq\lambda<\lambda_{c1}$. In the following, we
focus on the data obtained with $L=2048$.

Increasing $\lambda$ across each $\lambda_{c}$, $\sigma_{\rm SH}$
and $\sigma_{\rm CH}$ both exhibit sharp jumps: $\sigma_{\rm SH}$
jumps from $-4.30$ to $+4.87$  (in units of ${e/{8\pi}}$) at $\lambda_{c1}$,
from $+8.71$ to $+6.32$ at $\lambda_{c2}\approx0.56t$,
and from $+1.49$ to $-3.57$ at $\lambda_{c3}\approx0.71t$;
$\sigma_{\rm CH}$ changes as $+1\rightarrow-3\rightarrow+5\rightarrow+1$ (in units of $e^2/h$).
In intervals away from $\lambda_{c}$'s, $\sigma_{\rm SH}$ varies continuously while
$\sigma_{\rm CH}$ remains unchanged. The corresponding DOS [Fig.~\ref{f.1}(c-j)]
also points out that the lowest two Landau subbands
approach, merge together and form a pseudogap at each $\lambda_{c}$, and then separate
for three times.

Mainly, such a four-step evolution of the SHC of TBE is
distinct from the resonant SHC of 2DEG predicted by Shen {\it et
al.}~\cite{Shen} in four aspects: in 2DEG, two
Landau levels cross each other at the Fermi level only once and
produce one $\lambda_{c}$, while for TBE the two Landau
subbands touch successively three times and
results in three $\lambda_{c}$'s; at a $\lambda_{c}$, the SHC of 2DEG diverges at
$T=0$, while the SHC of TBE converges to finite values in
the thermodynamic limit ($L\rightarrow\infty$) at $T=0$; the
SHC of 2DEG does not change its sign while the SHC of TBE
changes its sign at $\lambda_{c1}$ and $\lambda_{c3}$;
furthermore, the CHC of 2DEG is unaffected when tuning $\lambda$,
but the CHC of TBE presents three successive quantum jumps.

\begin{figure}[!htb]
  \vspace{-0.25in}
  \hspace{-0.15in}
  \includegraphics[scale=0.5]{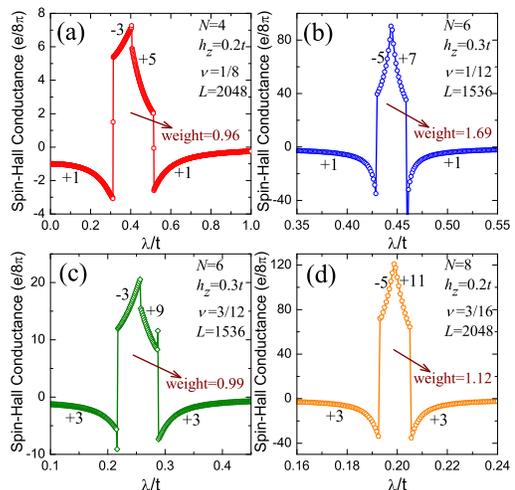}
  \vspace{-0.35in}
  \caption{(color online). $\sigma_{\rm SH}$ versus the Rashba SOC parameter $\lambda$ in
  various cases. $\sigma_{\rm CH}$ (in units of $e^2/h$) of each evolution step is also shown.} \label{f.2}
\end{figure}

{\it Cases with weaker magnetic fields.---}The above four-step
evolutions have
also been verified by further numerical calculations of the cases
with $N=4-16$, $h_{z}=0.05t-0.4t$, and various $\nu$'s (with odd
number of totally filled Landau subbands), as illustrated by four examples
in Fig.~\ref{f.2}. For $N=4$, $h_{z}=0.2t$ and $\nu=1/8$
[Fig.~\ref{f.2}(a)], $\sigma_{\rm SH}$ shows behaviors similar
to that in Fig.~\ref{f.1}(a), while with smaller $\lambda_{c}$'s
and narrower transition regions (i.e.,
smaller $\lambda_{c3}-\lambda_{c1}$);
for $N=6$ and $N=8$ [Fig.~\ref{f.2}(b-d)], the
transition regions are narrower than the case with $N=4$.
Meanwhile, the quantized CHC also exhibits three jumps
by $-Ne^2/h$, $+2Ne^2/h$ and $-Ne^2/h$.

In brief, the larger $N$'s, the significantly narrower are the
transition regions ($\lambda_{c1}\leq\lambda\leq\lambda_{c3}$).
However, the positive values in the transition regions are much
larger, and the total weights of positive part of $\sigma_{\rm SH}$
(i.e. the integral from $\lambda_{c1}$ to $\lambda_{c3}$) possessing
the same order of magnitude, are respectively $0.96$, $1.69$, $0.99$
and $1.12$ in the four cases of Fig.~\ref{f.2}. [Note that the weight is $1.22$ for
the case in Fig.~\ref{f.1}(a).]

\begin{figure}[!htb]
  \vspace{-0.2in}
  \hspace{-0.05in}
  \includegraphics[scale=0.4]{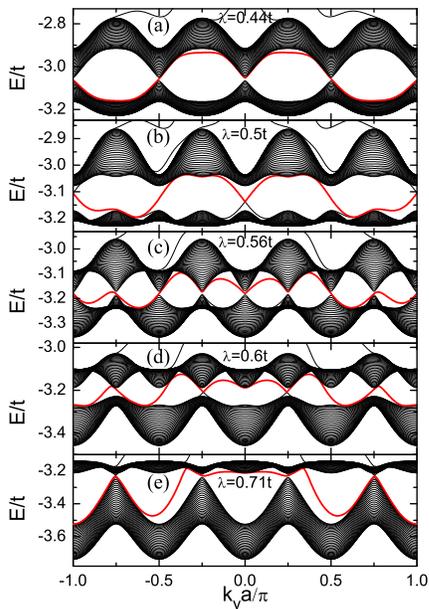}
  \vspace{-0.2in}
  \caption{(color online). Lowest two subbands and intermediate edge states
  [shown as thick (red) lines] of a cylinder of the size $128\times\infty$ (OBC in $x$ direction and PBC in $y$
  direction) with $N=4$, $h_{z}=0.4$ and various $\lambda$'s.} \label{f.3}
\end{figure}

{\it Edge states in a cylindrical system with $N=4$.---} An alternative way to
reveal the
distinctions among four evolution steps is to calculate the edge states of the system on a
cylinder. These edge states reflect the topological character of the
corresponding bulk state~\cite{Halperin,Hatsugai}. Just recently, spin-filtered edge states
have been considered for a graphene cylinder with an intrinsic SOC
~\cite{Kane} (in a two-component Haldane model~\cite{Haldane}) or
Zeeman splitting~\cite{Lee}, a quantum SHE arising from helical edge states
have been proposed and experimentally verified soon
in HgTe quantum wells~\cite{Bernevig}, and edge states have also
been employed to characterize topological band insulators and
chiral spin liquids~\cite{DHLee}. Now as an illustration,
we take a cylinder of square lattice of
the size $128\times\infty$ and apply open boundary condition (OBC)
in $x$ direction and PBC in $y$ direction.

Chern numbers of bulk Landau subbands are intimately related to
the winding numbers of the
corresponding edge states~\cite{Hatsugai}. For
$\lambda_{c1}<\lambda<\lambda_{c2}$, there is one edge state
winding three times from the upper subband to the lower one then back to
the upper one [a thick (red) line in Fig.~\ref{f.3}(b)] which corresponds
to a Chern number $-3$ of the lower subband. For
$\lambda_{c2}<\lambda<\lambda_{c3}$, there is one edge state
winding five times from the lower subband to the upper one then back to
the lower one [Fig.~\ref{f.3}(d)] which corresponds to a Chern number $+5$ of the lower subband.
While for $0<\lambda<\lambda_{c1}$ or $\lambda_{c3}<\lambda<1.0t$
(not shown in Fig.~\ref{f.3}), there is another edge state
winding only once from the lower subband to the upper one then back
to the lower one which corresponds
to a Chern number $+1$ of the lower subband.

The continuum spectrum of this cylinder also gives
further descriptions about the jumps of the bulk CHC. Increasing
$\lambda$ across $\lambda_{c1}$ [Fig.~\ref{f.3}(a)] or
$\lambda_{c3}$ [Fig.~\ref{f.3}(e)], the lowest two
subbands touch at four points simultaneously in $k$-space and a
Chern number $-4$ is transfered from the upper subband to the
lower one; while across $\lambda_{c2}$
[Fig.~\ref{f.3}(c)], the lowest two subbands touch at eight points
simultaneously in $k$-space and a Chern number $+8$ is
transferred between them. Such a correspondence between
transferred chern numbers and touching points in $k$-space has also been verified
for $N=5-8$.

\begin{figure}[!htb]
  \vspace{-0.2in}
  \hspace{-0.25in}
  \includegraphics[scale=0.6]{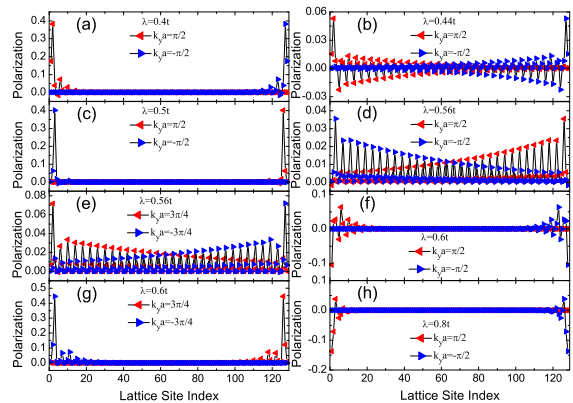}
  \vspace{-0.45in}
  \caption{(color online). Spin polarization $P^{z}$ (in units of ${\hbar/2}$) versus
  the lattice site index in $x$ direction for the edge states
  of the cylindrical system in Fig.~\ref{f.3}.}\label{f.4}
\end{figure}

In addition, the spin polarization carried by the edge states
can be computed explicitly as
$P^{z}_{m{\bf k}}(i)={\hbar/2}\langle m{\bf
k}|\hat{c}^{\dagger}_{i}\sigma_{z}\hat{c}_{i}|m{\bf k}\rangle$ with
$i$ the lattice site index in $x$ direction~\cite{Sheng3}.
In Fig.~\ref{f.4}, we plot the spin polarization $P^{z}$
of some edge states of the above cylindrical system.
If $\lambda$ takes a value far away from
$\lambda_{c}$'s, $P^{z}$ takes prominently
large values near the left or
the right edge and is almost zero in the intermediate region [Fig.~\ref{f.4}(a), (c) and (f-h)];
but if $\lambda$ takes
a value close to $\lambda_{c}$'s, $P^{z}$ fluctuates strongly between two edges
[Fig.~\ref{f.4}(b), (d) and (e)].
Note that for a fixed $k_{y}$, the dominantly
positive peak of $P^{z}$ moves to another edge when $\lambda$ varies
from $0.4t$ to $0.5t$. And for edge states of $\lambda=0.8t$
[Fig.~\ref{f.4}(h)], $P^{z}$ takes prominently negative values near edges.

{\it Presence of disorder.---} We add a term
$\sum_{i}w_{i}\hat{c}^{\dagger}_{i}\hat{c}_{i}$~\cite{Sheng2} into
the Hamiltonian [Eq.~(\ref{e.1})] to account for the effects of
nonmagnetic disorder, $w_i$ being a random potential uniformly distributed
between $[-W/2,W/2]$.

\begin{figure}[!htb]
  \vspace{-0.2in}
  \hspace{-0.18in}
  \includegraphics[scale=0.5]{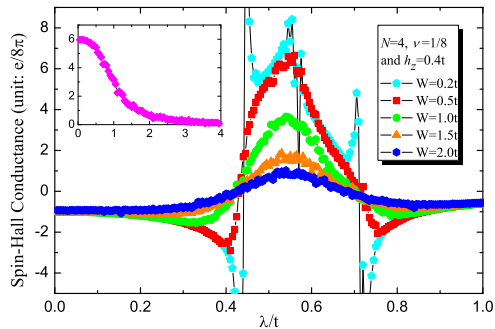}
  \vspace{-0.3in}
  \caption{(color online). $\sigma_{\rm SH}$ versus $\lambda$ in
  the case with $N=4$, $h_{z}=0.4t$ and various disorder strength
  $W$'s (100 random-potential configurations of the size $8\times8$). The inset shows the evolution of $\sigma_{\rm SH}$ versus $W$
at $\lambda=0.5t$.}\label{f.5}
\end{figure}

For $N=4$, the adopted $100$
random-potential configurations are of the size $8\times8$
(such a super unit cell is
commensurate with the magnetic unit cell in the absence of
disorder), and the total lattice is of the size $32\times32$. It can be seen
that weak disorder ($W\leq0.5t$) does not smear out the overall
four-step evolution of the SHC. For stronger disorder ($W=2.0t$),
the SHC does not show resonance anymore near $\lambda_{c1}$ or $\lambda_{c3}$,
and takes positive values in
an enlarged interval while the peak is diminished into a hump.

{\it Summary and discussion.---}An appealing evolution of
magneto-transport property has been demonstrated by TBE with
Rashba SOC in a magnetic field: (i) with the
flux strength $\phi={2\pi/N}$ and the Zeeman
splitting fixed, when increasing the Rashba SOC $\lambda$ from
$0$, four-step evolutions of
the SHC and CHC have been observed; (ii) at three $\lambda_{c}$'s, the
SHC shows size-dependent resonances and jumps, and changes its sign
at $\lambda_{c1}$ and $\lambda_{c3}$; (iii) meanwhile, the quantized CHC
shows three successive jumps by $-Ne^2/h$, $+2Ne^2/h$ and $-Ne^2/h$;
(iv) for smaller $\phi$'s, the total weights of
positive part of SHC have the same order of magnitude although the
transition regions are
significantly narrower; (v) edge states of a cylindrical system
reflect such bulk properties; (vi) this four-step evolution
is robust against weak disorder.

Such a four-step evolution of SHC is expected to occur
in 2D electron systems with a lattice potential,
a mechanism of SOC or SO scattering, and an external magnetic field.
Some candidate experimental systems are:
metallic conductors such as Al films and Pt strips~\cite{Valenzuela},
untracold fermions in an optical
lattice with an effective SOC~\cite{Stanescu}, and
 graphene with an intrinsic or Rashba SOC~\cite{Kane,Sheng3}.
And spin polarizations of edge states should be observable in
a four-terminal experimental setup~\cite{Kane,Lee}.

This work was supported by the National Nature Science Foundation of
China (No. 90503014), the State Key Program for Basic Researches of
China (No. 2006CB921802), China Postdoctoral Science Foundation (No.
20070410330) and Jiangsu Planned Projects for Postdoctoral Research
Funds (No. 0602021C).

\end{document}